\documentstyle{article}
\begin{document}
\begin{titlepage}
\title{Pasti-Sorokin-Tonin Actions in the Presence of Sources} 
\author{ R. Medina
\thanks{e-mail address: rmedina@power.ift.unesp.br} and
N. Berkovits \thanks{e-mail address: nberkovi@power.ift.unesp.br} 
\\ Insituto de F\'\i sica Te\'orica \\ 
Universidade Estadual Paulista \\
Rua Pamplona 145\\ 01405-900 S\~{a}o
Paulo, Brasil \\
IFT-P.030/97}
\maketitle

\begin{abstract}

Pasti, Sorokin and Tonin have recently constructed manifestly Lorentz-invariant
actions for 
self-dual field strengths and for Maxwell fields with manifest electromagnetic 
duality. Using the method of Deser, Gomberoff, Henneaux and
Teitelboim, we generalize these actions in the presence of sources. 

\end{abstract}

\end{titlepage}

\newpage

\section{Introduction}

$\frac{}{}$

Recent interest in duality has renewed the search for 
actions where duality symmetry and Lorentz invariance
are manifest. At the present time, there are two types of actions
where these symmetries are manifest. The first type of action is
quadratic but contains
an infinite number of fields. It was first discovered in two dimensions 
for describing chiral bosons \cite{McClain}, and later generalized to arbitrary
self-dual fields \cite{Devecchi} and to Maxwell fields with manifest
electromagnetic duality \cite{Martin}.
In ten dimensions, it was rediscovered by analyzing the
massless Ramond-Ramond sector of superstring field theory\cite{Nathan1}. By
studying the coupling of D-branes to the massless Ramond-Ramond string 
fields, this first type of action was generalized in the
presence of sources \cite{Nathan2}.

The second type of manifestly
covariant action with manifest duality contains a finite number of fields 
but is
non-polynomial in these fields \cite{Pasti1,Pasti2}. It was discovered by 
Pasti, Sorokin and
Tonin (PST) and ``covariantizes'' actions found earlier \cite{Deser1, Schwarz}
which 
were manifestly dual but not manifestly Lorentz-covariant. This PST action was
inspired by an incorrect action of Khoudeir and Pantoja 
\cite{Pantoja,Pasti23}. Equivalence among
different formulations is of interest not only at a classical level
\cite{Pasti1}, but also at a quantum level \cite{Rivelles}. Recently, 
Deser, Gomberoff, Henneaux, and Teitelboim showed how
to introduce sources into the non-manifestly covariant versions of these
actions \cite{Deser2}. Their method uses a field strength whose definition
is modified in the presence of sources, as well as 
a coupling term of the type ``$\vec{A} \cdot \vec{j}$''.

In this paper, we ``covariantize'' their procedure,
thereby generalizing the manifestly covariant 
PST actions
in the presence of sources.
In the second section of this paper, we generalize the PST action for 
self-dual fields \cite{Pasti1}, and in the
third section, we 
generalize the PST action for 
Maxwell fields with manifest electromagnetic duality \cite{Pasti2}.

Our conventions are as follows: We use uncapitalized latin letters to denote 
space-time indices and capitalized latin letters to denote space indices.
We work with the metric $g_{mn} = diag(-1,+1, \ldots, +1)$. 
The antisymmetrization of space-time indices of a 
tensor is done without adding any additional factor, for example,
\begin{eqnarray}
B_{[lm]n} = B_{lmn} - B_{mln}.
\end{eqnarray}
 
\section{Manifestly Lorentz-invariant chiral 2-form \\
action with sources}

$\frac{}{}$

In ref. \cite{Pasti1}, Pasti, Sorokin and Tonin write a manifestly Lorentz
invariant action for a chiral 2-form field, which is an antisymmetric field
$A_{mn}(x)$ propagating in a 
$5+1$ dimensional Minkowski space-time whose field strength
$F_{lmn}(x)$ is a self-dual field on shell.
This means that
\begin{eqnarray}
F_{lmn}(x) = *F_{lmn}(x),
\end{eqnarray} 
where, in general, the dual of a field $C_{lmn}(x)$ is defined as 
\begin{eqnarray}
*C_{lmn}(x) = \frac{1}{3!} \epsilon_{lmnpqr} C^{pqr}(x).
\end{eqnarray}
We are using the convention
\begin{eqnarray}
F_{lmn}(x) = \partial_l A_{mn}(x) + \partial_m A_{nl}(x) + 
\partial_n A_{lm}(x).
\label{def-F}
\end{eqnarray} 

The generalization of the action of ref. \cite{Pasti1} in the presence of 
sources is
\begin{eqnarray}
S=\int d^6x (-\frac{1}{6}H_{lmn}H^{lmn}
+\frac{1}{2(\partial_q a \ \partial^q a)} \partial^m a \ {\cal H}_{mnl}
{\cal H}^{nlr} \partial_r a - A_{ln} j^{ln}), 
\label{action}
\end{eqnarray}
where $j^{ln}$ is the source (for self-dual fields the electric and magnetic
sources are equal) and 
\begin{eqnarray}
H_{lmn} = F_{lmn} + *G_{lmn}
\end{eqnarray}
is the modified field strength. $F_{lmn}$ is still defined by (\ref{def-F})
and satisfies Bianchi identities $\partial^l(*F_{lmn})=0$.
$G_{lmn}$ is defined to satisfy
\begin{eqnarray}
\partial_l G^{lmn} - j^{mn} = 0.
\label{condition-G}
\end{eqnarray}

${\cal H}_{lmn}$ is an anti self-dual field defined as 
\begin{eqnarray}
{\cal H}_{lmn}= H_{lmn} - *H_{lmn}.
\label{anti}
\end{eqnarray}
The action in (\ref{action}) is manifestly Lorentz invariant and
coincides with the original action of ref. \cite{Pasti1} when 
there are no sources. We will show that it describes the dynamics of a chiral
2-form in the presence of sources by comparing this action, in a certain
gauge, with the non-manifestly covariant one of ref. \cite{Deser2}. 

Making use of the identity 
$u^{[l} {\cal H}^{mn]s} u_s = 2(u_s u^s) {\cal H}^{lmn} + 
\epsilon^{lmnsuv} u^w u_s {\cal H}_{uvw}$ for $u_l = \partial_l a$,
the equation of motion for $A_{mn}(x)$ leads to
\begin{eqnarray}
\epsilon^{lmnpqr} \partial_n \left( \frac{1}{(\partial_t a \ \partial^t a)} 
\partial_p a \ {\cal H}_{qrs} \ \partial^s a \right) = 0.
\label{eq-motion}
\end{eqnarray}
And now, considering the identity
$$\epsilon^{lmnpqr} \partial_n (T_{lm} \ \partial_p a \ T_{qr}) =
2 \ T_{lm} \ \epsilon^{lmnpqr} \partial_n 
(\partial_p a \ T_{qr}),$$
in the case of
$T_{lm} = \frac{1}{(\partial_k a \ \partial^k a)} {\cal H}_{lmi}\partial^i a$,
the equation of motion for $a(x)$ may be written as
\begin{eqnarray}
\left \{ \frac{1}{(\partial_k a \ \partial^k a)} {\cal H}_{lmi} 
\partial^i a \right \} \ \left \{ \epsilon^{lmnpqr} \partial_n \left(
\frac{1}{(\partial_t a \ \partial^t a)} \partial_p a \ {\cal  H}_{qrs}      
\partial^s a \right) \right \} = 0.    
\label{eq2-motion}
\end{eqnarray}
Thus, when considering (\ref{eq-motion}), this last equation becomes simply
an identity. This happens because $a(x)$ is a gauge field as can be
seen by observing that the following gauge transformation leaves the action
invariant:    
\begin{eqnarray}
\delta a(x) = \phi (x),\ \ \
\delta A_{mn} (x) = \frac{\phi(x)}{(\partial_q a \ \partial^q a)}
{\cal H}_{mns} \ \partial^s a. 
\label{gauge2}
\end{eqnarray}

Since $a(x)$ transforms without derivatives, its gauge choice can be directly
substituted in the action (\ref{action}) leading to a Lorentz non-covariant
formulation of it. Choosing the gauge 
$\partial_m a = \delta^0_m$, this substitution leads to
\begin{eqnarray}
S=\int d^6x (-\frac{1}{6}H_{lmn}H^{lmn}
+\frac{1}{2} {\cal H}_{0nl}
{\cal H}^{nl0} - A_{ln} j^{ln}). 
\label{action3}
\end{eqnarray}
Using that $H_{lmn}H^{lmn}= 3 H_{0AB}H^{0AB} - 3 *H_{0AB} *H^{0AB}$ and
identifying $E^{AB}=-H^{0AB}$ and $B^{AB}=-*H^{0AB}$ we end up with
\begin{eqnarray}
S=\int d^6x ( E^{AB}B_{AB} - B^{AB}B_{AB} - A_{ln} j^{ln}), 
\label{action4}
\end{eqnarray}
which, except for a factor $\frac{1}{4}$, is the action of Deser, Gomberoff,
Henneaux and Teitelboim \cite{Deser2} \footnote{ In deriving (\ref{action4})
we have made the following identification between the 5 index Levi-Cita symbol 
used in ref. \cite{Deser2} and ours: $\epsilon^{ABCDE}=\epsilon^{ABCED0}$. }.
 
\section{Lorentz-covariant formulation for duality \\ 
symmetric Maxwell action with sources}

$\frac{}{}$

In this section we generalize the PST action for Maxwell fields \cite{Pasti2} 
by considering the coupling of these fields to external sources. Their action
is a Lorentz-invariant version of the one found by Schwarz and Sen
\cite{Schwarz} 
in which, for the purpose of having manifest duality in the
action, an additional Abelian field is included. 

The procedure
is straightforward from what we did in the previous section. In this case the
action is
\begin{eqnarray}
S = \int d^4 x \left [-\frac{1}{8} H^{\alpha}_{mn}
 H_{\alpha}^{mn}
- \frac{1}{4(\partial_q a \ \partial^q a)} \partial^m a \ 
{\cal H}^{\alpha}_{mn} {\cal H}_{\alpha}^{np} \ \partial_p a + 
\frac{1}{2} A^{\alpha}_m \epsilon_{\alpha \beta} j^{\beta m}  \right ] ,
\label{action2}
\end{eqnarray}
where the two Abelian fields are $A^{\alpha}_n$ $(\alpha = 1, 2)$ and
\begin{eqnarray}
\label{F}
F^{\alpha}_{mn} = \partial_m A^{\alpha}_n - \partial_n A^{\alpha}_m, \\
\label{H}
H^{\alpha}_{mn} = F^{\alpha}_{mn} + * G^{\alpha}_{mn}, \\
\label{dual}
{\cal H}^{\alpha}_{mn} = \epsilon^{\alpha \beta} H_{\beta mn} - 
*H^{\alpha}_{mn}.
\end{eqnarray}
$\epsilon^{\alpha \beta}$ is completely antisymmetric on its indices, with 
$\epsilon^{12}= 1$. The metric tensor for the internal space is
$g_{\alpha \beta} = \delta_{\alpha \beta}$.

Again, $G^{\alpha mn}$ is an external field, related to the sources by the
condition
\begin{eqnarray}
\partial_m G^{\alpha mn} + j^{\alpha n} = 0.
\label{sources2}
\end{eqnarray}

Our action (\ref{action2}) coincides with that of Pasti, Sorokin and
Tonin \cite{Pasti2} when there are no sources.
Note that the field ${\cal H}^{\alpha}_{mn}$ defined in (\ref{H}) is 
self-dual with respect to the Lorentz and internal indices: 
${\cal H}^{\alpha}_{mn} = \frac{1}{2} \epsilon^{\alpha \beta}
\epsilon_{mnpq} {\cal H}_{\beta}^{pq}$. 

Now, using the identity $u_p {\cal H}^{\alpha pn} u^{m} = 
- (u^p u_p) {\cal H}^{\alpha nm} - \epsilon^{nmst} \epsilon^{\alpha \beta}
u^q {\cal H}_{\beta qs} u_t + u_p {\cal H}^{\alpha pm} u^{n}$ for 
$u_l = \partial_l a$, the equations of motion with respect to
$A_{\alpha n}(x)$ and $a(x)$ lead respectively to
\begin{eqnarray}
\epsilon^{mnpq} \partial_n \left[ \frac{1}{(\partial_s a \partial^s a)} 
\partial_p a \ {\cal H}^{\alpha}_{qr} \ \partial^r a \right ] = 0,
\label{eq4-motion}
\end{eqnarray}
\begin{eqnarray}
\left \{ \epsilon_{\beta \alpha} \frac{ \partial^l a \ {\cal H}^{\beta}_{ml} }
{ (\partial_k a \ \partial^k a) } \right \}
\left \{ \epsilon^{mnpq} \partial_n 
\left[ \frac{1}{(\partial_s a \partial^s a)} 
\partial_p a \ {\cal H}^{\alpha}_{qr} \ \partial^r a \right ] \right \} = 0.
\label{eq5-motion}
\end{eqnarray}
Again, equation (\ref{eq5-motion}) becomes an identity when (\ref{eq4-motion})
is considered and this happens because $a(x)$ is a gauge field.
This can be seen by noting that the following gauge transformation leaves the 
action (\ref{action2}) invariant:
\begin{eqnarray}
\delta a(x) = \phi (x) , \ \ \ 
\delta A^{\alpha}_m (x) = \phi (x) \frac{1}{(\partial_q a \ \partial^q a)}
\epsilon^{\alpha \beta} {\cal H}_{\beta mn} \partial^n a.
\end{eqnarray}
In the same way as was noted in the previous section, as the transformation for
$a(x)$ doesn't involve any derivatives, the gauge fixing of this field can be
done at the level of the action. Using the gauge $\partial_m a = \delta^0_m$
and the identity
$$H^{\alpha}_{mn} H_{\alpha}^{mn} = 2 H^{\alpha}_{0N} H_{\alpha}^{0N} 
- 2 *H_{0N} *H^{0N} \ (N=1,2,3),$$ 
the action (\ref{action2}) adopts the form
\begin{eqnarray}
S = \frac{1}{2} \int d^4 x \left [ \epsilon_{\alpha \beta} \vec{B}^{\alpha} 
\cdot \vec{E'}^{\beta} - \vec{B}^{\alpha} \cdot \vec{B}_{\alpha} +
A^{\alpha}_m \epsilon_{\alpha \beta} j^{\beta m}  \right ] ,
\label{action5}
\end{eqnarray}
where ${E'}^{\alpha N} = H^{\alpha 0N}$ and $B^{\alpha N}=*H^{\alpha ON}$. In 
(\ref{action5}) we have 
${E'}^{\alpha N} = E^{\alpha N} - \partial^N A^{\alpha 0}$, where 
$E^{\alpha N}$ is the ``electric field'' appearing in \cite{Deser2}.
The action (\ref{action5}) coincides with the Maxwell action of
ref. \cite{Deser2}, in the case of external sources, after cancelling
the term
$A^{\alpha}_0 \epsilon_{\alpha \beta} j^{\beta 0}$ with the contribution coming
from $\partial^N A^{\alpha 0}$ in ${E'}^{\alpha N}$.

\section{Acknowledgements}

$\frac{}{}$

RM ackowledges support from a FAPESP grant No. 96/00702-8. NB would like to
acknowledge the financial support of FAPESP grant No. 96/05524-0 and useful
conversations with S. Deser and D. Sorokin.

\end{document}